\documentclass[12pt]{article}
\usepackage{amssymb,amsmath,latexsym}
\begin{document}
\renewcommand{\emptyset}{\varnothing}
\newcommand{\UU}{{\cal U}} 
\newcommand{\Sch}{{\cal S}} 
\newcommand{\TA}{{\cal T}} 
\newcommand{\inst}{{\cal I}} 
\newcommand{\tup}{{\bf t}} 
\newcommand{\DB}{{\bf D}} 
\newcommand{\relvars}{{\it Relvars}}
\newcommand{\typevars}{{\it Typevars}}
\newcommand{\specattrs}{{\it Specattrs}}
\newcommand{\constraint}{{\it constraint}}
\newcommand{\decl}{{\it decl}}
\newcommand{\outvars}{{\it Outvars}}
\newcommand{\outatt}{{\it outatt}}
\newcommand{\lsem}{\mathopen{\hbox{$[\![$}}}
\newcommand{\rsem}{\mathclose{\hbox{$]\!]$}}}
\newcommand{\sem}[1]{\lsem #1 \rsem}
\newcommand{\semijoin}{\ltimes}

\newtheorem{theor}{Theorem}
\newtheorem{propo}{Proposition}
\newtheorem{defin}{Definition}
\newtheorem{corol}{Corollary}
\newtheorem{lemma}{Lemma}

\newlength{\qedlengte}
\settowidth{\qedlengte}{$\Box$}
\addtolength{\qedlengte}{-0.25\qedlengte}
\newcommand{\qedbox}{\rule{\qedlengte}{\qedlengte}}
\newcommand{\qed}{\hspace*{1em}\hfill\qedbox}

\newcommand{\vergeet}[1]{}

\title{Polymorphic type inference \\ for the relational algebra}
\author{\textsl{Jan Van den Bussche} \\ Limburgs Universitair Centrum \\
Belgium \\
jan.vandenbussche@luc.ac.be \and
\textsl{Emmanuel Waller} \\ LRI, Universit\'e Paris Sud \\ France \\
emmanuel.waller@lri.fr}
\date{}
\maketitle

\begin{abstract}

We give a polymorphic account of the relational algebra. We introduce a
formalism of ``type formulas'' specifically tuned for relational algebra
expressions, and present an algorithm that computes the ``principal''
type for a given expression. The principal type of an expression is a
formula that specifies, in a clear and concise manner, all assignments
of types (sets of attributes) to relation names, under which a given
relational algebra expression is well-typed, as well as the output type
that expression will have under each of these assignments.  Topics
discussed include complexity and polymorphic expressive power.

\end{abstract}

\section{Introduction}

The operators of the relational algebra (the basis of all relational
query languages) are polymorphic.  We can take the natural join of any
two relations, regardless of their sets of attributes. We can take the
union of any two relations over the same set of attributes. We can take
the cartesian product of any two relations having no attributes in
common. We can perform a selection $\sigma_{A<B}$ on any relation having
at least the attributes $A$ and $B$.  Similar typing conditions can be
formulated for the other operators of the relational algebra.  When
combining operators into expressions, these typing conditions can become
more involved.  For example, for the expression $$ \sigma_{A<5}(r \Join
s) \Join ((r \times u) - v) $$ to be well-typed, the attribute $A$ must
be an attribute of $r$ or $s$ (or both). But if it is an attribute of
$r$, then it must also be one of $v$. Moreover,
by the subexpression $(r\times u)-v$, the relation schemas
of $r$ and $s$ must be disjoint, and their union must be the type of $v$.

A natural question thus arises: given a relational algebra expression $e$,
under which database schemas is $e$
well-typed?  And what is the result relation schema of $e$
under each of these assignments?
This is nothing but the relational algebra version of the
classical {\em type inference\/} problem.
Type inference is an extensively studied topic in the theory of
programming languages
\cite{asu,hindley,mitchell,tiuryn_survey,giannini_survey}, and is used in
industrial-strength functional programming languages such as SML/NJ
\cite{ullman_ml}.

Doing type inference for some language involves setting up two things.
First, we need a system of
{\em type rules\/} that allow to derive the output type of a program
given types for its input parameters.  Typically such an output type can only
be derived for some of all possible assignments of types to input parameters;
under these assignments the program is said to be {\em well-typed.}
Second, we need a formalism of {\em type formulas.} A type formula
defines a family of input type assignments, as well as an output
type for each type assignment in the family.
Every typable program should have a {\em principal\/} type formula, which
defines all type assignments under which the program is well-typed,
as well as the output type of the program under each of these assignments.
The task then is to come up with a
{\em type inference\/} algorithm that will compute the principal type for
any given program.

In this paper, we do type inference for the relational algebra.  The
relational algebra is very different from the programming languages usually
considered in type inference; two fundamental features of such languages,
higher-order functions and data constructors (function symbols) are completely
absent here.  On the other hand, the set-based nature of relation types, and
the particulars of the standard relational algebra operators when viewed
polymorphically, present new challenges.  As a
consequence, our formalism of type formulas is drastically different from the
formalisms used in the theory of programming languages.

Our main motivation for this work was foundational and theoretical; after all,
query languages are specialized programming languages, so important ideas from
programming languages should be applied and adapted to the query language
context as much as possible.  However,
we also believe that type inference for database
query languages is tied to the familiar
principle of ``logical data independence.''  By this principle, a query
formulated on the logical level must not only be insensitive to changes on the
physical level, but also to changes to the database schema, as long as
these changes are to parts of the schema on which the query does not
depend.  To give a trivial example, the SQL query {\tt select * from R
where A<5} still works if we drop from {\tt R} some column
{\tt B} different from {\tt A}, but not if we drop column {\tt A}
itself. Turning this around, it
is thus useful to infer, given a query, under exactly
which schemas it works, so that the programmer sees to which schema
changes the query is sensitive.

Some recent trends in database systems seem to add weight to the above
motivation.  {\em Stored procedures\/} \cite{melton} are 4GL and SQL code
fragments stored in database dictionary tables. Whenever the schema
changes, some of the stored procedures may become ill-typed, while
others that were ill-typed may become well-typed. Knowing the principal
type of each stored procedure may be helpful in this regard.  Models of
{\em semi-structured\/} data \cite{tsimmis,unql} loosen (or completely
abandon) the assumption of a given fixed schema.  Query languages for these
models are essentially schema-independent.
Nevertheless, as argued by Buneman et al.\
\cite{buneman_structure},
querying is more effective if at least some form of schema is available,
computed from the particular instance.
Type inference can be helpful in telling for which schemas a given query
is suitable.

Ohori, Buneman and Breazu-Tannen were probably the first to introduce
type inference in the context of database programming languages,
in their work on the language Machiavelli \cite{machiavellic,machiavellij}.
Machiavelli features polymorphic
field selection from nested records, as well as a polymorphic join
operator.  However, the inference of principal types for full-fledged
relational algebra expressions was not taken up in that work.
We should also mention the work of Stemple et al.\ \cite{stemple_poly}, who
investigated reflective implementations of the polymorphic relational algebra
operators.

Other important related work is that on the extension of functional
programming languages with polymorphic record types.  Some of
the most sophisticated proposals in that direction were made by
R\'emy \cite{remy_records,remy_concatenation}.
This work adds record types to the type system of ML, featuring polymorphic
field selection and record concatenation.  While this system captures
many realistic functional programs involving records, it cannot
express the conditions on the types of relations implied by
certain relational algebra expressions, such as the example we gave
earlier.  Notably constraints
such as set disjointness (needed for the operator $\times$)
or set equality (for the operator $\cup$), cannot be expressed 
in other systems.  The reason is probably the additional concern of
these systems for subtyping: a program applicable to records of a certain type
should more generally
be applicable to records having all the fields of that type and
possibly more.  This is clearly not true for relational algebra expressions.

If one is only interested in deciding whether a given relational
algebra expression is {\em typable\/} (i.e., whether there exists at
least one schema under which the expression is well-typed), we show
that this problem is in the complexity class NP\@.  

In a final section of this paper, we formally
define the notion of {\em polymorphic query}.
Using our type inference algorithm, we prove that various operators
usually considered ``derived,'' because they can be simulated using the
standard relational algebra operators (e.g., semijoin), can {\em not\/} be
simulated {\em in a polymorphic way}.
Thus, our work also brings up new issues in the design of appropriate
polymorphic query languages.  

\section{Preliminaries}\label{prelims}

\subsection{Schemas, types, and expressions}
Assume given sufficiently large supplies of {\em relation variables\/} and
of {\em attribute names}.
Relation variables will be denoted by lowercase
letters from the end of the alphabet. Attribute names
will be denoted by uppercase letters from the beginning of the alphabet.

A {\em schema\/} is a finite set $\Sch$ of relation variables.
A {\em type\/} is a finite set $\tau$ of attribute names.  
Let $\Sch$ be a schema.  A {\em type assignment on
$\Sch$\/} is a mapping $\TA$ on $\Sch$,
assigning to each $r \in \Sch$ a type $\TA(r)$.  So, we have split the usual
notion of database schema, which specifies both the relation names and the
associated sets of attributes, in two notions.

The expressions of the {\em relational algebra\/} are defined
by the following grammar:
\begin{eqnarray*}
e & \to & r \\
& \mid & (e \cup e) \mid (e - e) \mid (e \Join e) \mid (e \times e) \\
& \mid & \sigma_{\theta(A_1,\ldots,A_n)}(e)
\mid \pi_{A_1,\ldots,A_n}(e) \mid \rho_{A/B}(e) \mid
\widehat{\pi}_A(e)
\end{eqnarray*}
Here $e$ denotes an expression,
$r$ denotes a relation variable, and
$A$, $B$, and $A_i$ denote attribute names.  The $\theta$ denotes a selection
predicate.

The schema consisting of all relation variables
occurring in expression $e$ is denoted by $\relvars(e)$.

\subsection{Well-typed expressions}
Let $\Sch$ be a schema,
$e$ an expression with $\relvars(e) \subseteq \Sch$, $\TA$ a
type assignment on $\Sch$, and $\tau$ a type.
The rules for when {\em $e$ has
type $\tau$ given $\TA$}, denoted by $\TA \vdash e:\tau$,
are the following:

\begin{center}
\newcommand{\plushaut}{\rule{0pt}{4ex}}
$ \displaystyle {\TA(r) = \tau \over \TA \vdash r:\tau} \plushaut $
\qquad
$ \displaystyle {\TA \vdash e_1:\tau \quad \TA \vdash e_2:\tau \over
\TA \vdash (e_1 \cup e_2) : \tau} \plushaut $
\qquad
$ \displaystyle {\TA \vdash e_1:\tau \quad \TA \vdash e_2:\tau \over
\TA \vdash (e_1 - e_2) : \tau} \plushaut $
\qquad
$ \displaystyle {\TA \vdash e_1:\tau_1 \quad \TA \vdash e_2:\tau_2 \over
\TA \vdash (e_1 \Join e_2) : \tau_1 \cup \tau_2} \plushaut $
\qquad
$ \displaystyle {\TA \vdash e_1:\tau_1 \quad \TA \vdash e_2:\tau_2 \quad
\tau_1 \cap \tau_2 = \emptyset \over
\TA \vdash (e_1 \times e_2) : \tau_1 \cup \tau_2} \plushaut $
\qquad
$ \displaystyle {\TA \vdash e:\tau \quad A_1,\ldots,A_n \in \tau \over
\TA \vdash \sigma_{\theta(A_1,\ldots,A_n)}(e):\tau} \plushaut $
\qquad
$ \displaystyle {\TA \vdash e:\tau \quad A_1,\ldots,A_n \in \tau \over
\TA \vdash \pi_{A_1,\ldots,A_n}(e):\{A_1,\ldots,A_n\} } \plushaut $
\qquad
$ \displaystyle {\TA \vdash e:\tau \quad A \in \tau \quad B \not \in \tau
\over \TA \vdash \rho_{A/B}(e):(\tau - \{A\}) \cup \{B\} } \plushaut $
\qquad
$ \displaystyle {\TA \vdash e:\tau \quad A \in \tau
\over \TA \vdash \widehat \pi_A(e):\tau - \{A\} } \plushaut $
\end{center}

We have a first basic definition:
\begin{defin} \rm
Let $e$ be an expression and let $\TA$ be a type assignment on $\relvars(e)$.
If there exists a type $\tau$ such that $\TA \vdash e:\tau$,
we say that $e$ is {\em well-typed under $\TA$.}
\end{defin}
Note that in this case $\tau$ is
unique and can easily be derived from $\TA$ by applying the rules in an order 
determined by the syntax of the expression $e$.

\subsection{Semantics}
We assume given a universe $\bf U$ of {\em data elements}.

Let $\tau$ be a type. A {\em tuple of type $\tau$} is a mapping
$\tup$ on $\tau$,
assigning to each $A \in \tau$ a data element $\tup(A) \in {\bf U}$.
A {\em relation of type $\tau$} is a finite set of tuples of type $\tau$.

Let $\Sch$ be a schema, and let $\TA$ be a type assignment on $\Sch$.
A {\em database of type $\TA$}
is a mapping $\DB$ on $\Sch$, assigning to each $r \in
\Sch$ a relation $\DB(r)$ of type $\TA(r)$.

The semantics of well-typed relational algebra expressions is 
the well-known one.  If $\TA \vdash e:\tau$,
and $\DB$ is a
database of type $\TA$, then the {\em result of evaluating $e$ on $\DB$}
is a relation of type $\tau$ defined in the well-known manner.
The only operator worth mentioning is perhaps the not so usual
$\widehat \pi_A$, which projects
{\em out\/} the attribute $A$, leaving all others intact.

At this point a remark is in order concerning the non-redundancy of the set of
relational operators we consider.  We have included both the natural join
$\Join$ and the cartesian product $\times$, and also both the standard
projection $\pi_{A_1,\ldots,A_n}$ and the ``complementary''
projection $\widehat \pi_A$.
It is well known that if
the type assignment is fixed and known, $\Join$ can be simulated using
$\times$ (plus selection and renaming), and conversely,
$\times$ can be simulated
using $\Join$ (plus renaming).
Also, $\pi$ can be simulated by a series of
$\widehat \pi$'s, and $\widehat \pi$ can be simulated by $\pi$.
To illustrate
the latter, if we fix the type of $r$ to $\{A,B,C\}$, then $\pi_A(r)$ is
equivalent to $\widehat \pi_B \widehat \pi_C (r)$, and $\widehat \pi_A(r)$ is
equivalent to $\pi_{B,C}(r)$.  However, these simulations are not
``polymorphic,'' in the sense that they depend on the particular type
assignment.

As a matter of fact,
we will see in Proposition~\ref{nonredundant} that
{\em polymorphic\/} simulations of $\Join$ using
$\times$, or vice versa, and of $\pi$ using $\widehat \pi$, or vice versa,
do not exist.  Hence, from a polymorphic point of view, our chosen
set of relational algebra operators is {\em non-redundant.}

\section{Typable expressions}

The central notion of this paper is defined as follows:
\begin{defin} \rm
Expression $e$ is called {\em typable\/} if there 
exists a type assignment $\TA$ on $\relvars(e)$
such that $e$ is well-typed under $\TA$.
\end{defin}
A very simple example of an expression that is not typable is
$\sigma_{A=B}(\pi_{B,C} (r))$.

Is typability a decidable property?  This question is easily answered
by the following lemma.  We use the following notation. If
$\cal T$ is a type assignment and $\cal A$ is a set
of attribute names, then we denote by ${\cal T}|_{\cal A}$ the type assignment
defined by ${\cal T}|_{\cal A}(r) := {\cal T}(r) \cap {\cal A}$.
If $e$ is an expression then we denote the set of all attribute names that
explicitly occur in $e$ by ${\it Specattrs}(e)$.

\begin{lemma}
If ${\cal T} \vdash e : \tau$ and ${\cal A} \supseteq {\it Specattrs}(e)$,
then ${\cal T}|_{\cal A} \vdash e : \tau \cap {\cal A}$.
\end{lemma}

The proof is straightforward.
As a consequence, in order to decide whether there
exists a type assignment under which $e$ is well-typed, it suffices to
consider type assignments $\cal T$ with the property that ${\cal T}(r)
\subseteq {\it Specattrs}(e)$ for every $r$.  It follows immediately
that typability is in NP\@.  Whether or not it is in P, or is NP-complete,
remains open.

Of course, we are not satisfied simply by knowing whether or not
a given expression is typable.  What we really want is
a clear, concise picture of exactly under which type assignments
it is well-typed, as well as of
what type the expression will have under each of
these type assignments.  (Note that there
will in general be infinitely many such type assignments.)

In the following, we will define the formalism of {\em type formulas}, which
is specifically tuned towards this task.

\section{Examples of type formulas} \label{secexamples}

Consider the expression $$ e = \sigma_{B=C} ( ( \rho_{A/B}(r) \cup
s ) \Join u ). $$ This expression is well-typed under
exactly those
type assignments $\TA$ satisfying the following two conditions:
\begin{enumerate}
\item
$\TA(s) = (\TA(r) - \{A\}) \cup \{B\}$;
\item
$C$ must belong to at least one of $\TA(u)$, $\TA(r)$, or $\TA(s)$.
\end{enumerate}
Given such a $\TA$, the type of $e$ then will equal $\TA(s) \cup \TA(u)$.

All the above information is expressed by the following
type formula for $e$:
$$ \begin{array}{lcl}
\begin{array}{l}
r:a_1a_2 \\
s:a_1a_2 \\
u:a_2a_3
\end{array}
& \mapsto &
\begin{array}{l}
e:a_1a_2a_3
\end{array}
\\
A:r \land \neg s && A:u \\
B:s \land \neg r && B:{\bf true} \\
C:(r \leftrightarrow s) \land (r \lor s \lor u) && C:{\bf true}
\end{array} $$
This type formula will the output of our type inference algorithm.  It can be
intuitively read as follows.  {\em
Expression $e$ is well-typed under precisely all type assignments that can be
produced by the following procedure:
\begin{enumerate}
\item
Instantiate $a_1$, $a_2$ and $a_3$ by any three types, on condition that they
are pairwise disjoint, and do not contain $A$, nor $B$, nor $C$.
\item
{\em Preliminarily\/} assign type $a_1 \cup a_2$ to $r$; $a_1 \cup a_2$ to $s$;
and $a_2 \cup a_3$ to $u$.
\item
In this preliminary type assignment, $A$ must be
added to the type of $r$, but must not be added to that of $s$; whether it is
added to the type of $u$ is a free choice.
\item
Similarly, $B$ must be added to
the type of $s$, not to that of $r$, and freely to that of $u$.
\item
Finally, $C$ must be added at least to one of the types of $r$, $s$, and $u$,
but if we add it to $r$ we must also add it to $s$ and vice versa.
\end{enumerate}
The type of $e$ under a type assignment thus produced equals $a_1 \cup a_2
\cup a_3$, to which we must add $B$ and $C$, and to which we also add $A$ on
condition that it belongs to the type of $u$.}

The symbols $a_1$, $a_2$ and $a_3$ are called {\em type variables}.
The attributes $A$, $B$ and $C$, which are explicitly mentioned by the
expression, are called the {\em special attributes\/}
of the expression.  The {\em declaration\/}
of each relation variable as a string
of type variables (where concatenation denotes union) provides the {\em
polymorphic basis\/} of the type assignments under which the expression is
well-typed.  An {\em attribute constraint\/} for each special attribute
then specifies (by a Boolean formula) the allowed extensions of the
polymorphic basis types with that attribute.  The declarations and constraints
together form the {\em type context\/}; this is the left-hand side of the type
formula.  On the right-hand side we find the polymorphic basis of the
output type, and again for each special attribute, an {\em output condition\/}
which specifies (by a Boolean formula) under which condition that attribute
has to be added to the output type.

Let us see two more examples.
The type formula for the expression $$ e=
\pi_A(r) - \pi_A((\pi_A(r) \times s) - r), $$ which the reader will recognize
as the textbook expression for the division operator, is:
$$ \begin{array}{lcl} \begin{array}{l} r:a \\ s:a \end{array}  & \mapsto &
e:\emptyset \\
A:r \land \neg s && A:{\bf true} \end{array} $$
So $r$ and $s$ must have the same type except that $r$ has an additional $A$
(which $s$ has not).  The output type is always $\{A\}$.

The type formula for the expression discussed in the Introduction,
$$ e = \sigma_{A<5}(r \Join s) \Join ((r \times u) - v), $$ is:
$$ \begin{array}{@{}lcl}
\begin{array}{l} v:a_1a_2a_3a_4 \\ r:a_1a_3 \\ u:a_2a_4
\\ s:a_3a_4a_5 \end{array} & \mapsto & e:a_1a_2a_3a_4a_5 \\
A:(r\lor s) \land (v \leftrightarrow (r \lor u))
\land \neg(r \land u) && A:{\bf true} 
\end{array} $$
The declarations specify exactly, in a manner similar to Venn diagrams,
the conditions required on the types of the
relation variables for the expression to be well-typed.

\section{Type formulas and type inference --- Formal definitions}

Before we can describe our type inference algorithm, we need precise
definitions of the underlying formalism.  In what follows,
we assume given a sufficiently large supply of 
{\em type variables}.

\subsection{Type contexts}
A {\em type context\/} is a structure consisting of the following
components:
\begin{enumerate}
\item
A finite set $\relvars$ of relation variables.
\item
A finite set $\typevars $ of type variables.
\item
A mapping $\decl$ from $\relvars$ to $2^{\typevars}$,
called the {\em declaration mapping}.
\item
A finite set $\specattrs $ of attribute names (called the {\em special
attributes\/}).
\item
A mapping $\constraint$ on $\specattrs$, assigning to
each special attribute a Boolean formula over $\relvars$.
\end{enumerate}
We will usually denote a type context by the letter $\Gamma$ and, when
necessary to avoid ambiguities, will write $\relvars(\Gamma)$,
$\typevars(\Gamma)$, etc.

\subsection{Semantics of type contexts} 
Fix a type context $\Gamma$.
The ``models'' of $\Gamma$
will be type assignments on $\relvars(\Gamma)$.  We go from type contexts to
type assignments via the notion of instantiation.  An {\em instantiation of
$\Gamma$} is a mapping $\inst$ on $\typevars \cup \specattrs$,
such that
\begin{enumerate}
\item
$\inst$ assigns to each type variable a type, such that
\begin{itemize}
\item
for different type variables $a_1$ and $a_2$, $\inst(a_1)$ and $\inst(a_2)$
are disjoint; and
\item
for each type variable $a$ and special attribute $A$;
$A \not \in \inst(a)$.
\end{itemize}
\item
$\inst$ assigns to each special attribute a subset of $\relvars$, such that
for each special attribute $A$, $\inst(A) \models
\constraint(A)$.
(Since
$\constraint(A)$ is a Boolean formula over $\relvars$, and
$\inst(A)$ is a subset of $\relvars$,
the meaning of
$\inst(A) \models \constraint(A)$ is the standard meaning from propositional
logic.)
\end{enumerate}
If some of the Boolean formulas in $\Gamma$ are unsatisfiable, we call also
$\Gamma$ {\em unsatisfiable}.  In this case, $\Gamma$ has no
instantiations.

From a type context $\Gamma$ and an instantiation $\inst$ of $\Gamma$, we can
uniquely determine a type assignment $\TA$ on $\relvars$, defined on each
relation variable $r$ as follows:
$$ \TA(r) := \bigcup \{ \inst(a) \mid a \in \decl(r) \}
\cup \{A \in \specattrs \mid r \in \inst(A)\}. $$
We call this type assignment $\TA$ the {\em image of $\Gamma$ under $\inst$},
and conveniently denote it by $\inst(\Gamma)$.

\subsection{Type formulas}
A {\em type formula\/} now is a quadruple
$ (\Gamma, e, \outvars, \outatt)$,
where
\begin{enumerate}
\item
$\Gamma$ is a type context;
\item
$e$ is a relational algebra expression with $\relvars(e) \allowbreak
= \allowbreak \relvars(\Gamma)$, and such that
$\specattrs(\Gamma)$ contains all the attribute names that are explicitly
mentioned in $e$.
\item
$\outvars$ is a subset of $\typevars(\Gamma)$; and
\item
$\outatt$ is a mapping on $\specattrs(\Gamma)$, assigning to each special
attribute a Boolean formula over $\relvars(\Gamma)$.
\end{enumerate}

The way we write down concrete instances of type formulas has already been
illustrated in Section~\ref{secexamples}.

\subsection{Semantics of type formulas}
From a type formula $(\Gamma,\allowbreak e,\allowbreak \outvars,\allowbreak
\outatt)$
and an instantiation $\inst$
of $\Gamma$, we can uniquely determine the following type:
$$ \{\inst(a) \mid a \in \outvars\}
\cup \{A \in \specattrs \mid \inst(A) \models \outatt(A)\}. $$
We call this type the {\em output type of the type formula under $\inst$.}

We are now ready to define the following fundamental property of type
formulas:
\begin{defin} \rm
A type formula
$(\Gamma,\allowbreak e,\allowbreak \outvars,\allowbreak \outatt)$ is called
{\em principal for $e$\/}
if for every type assignment $\TA$ on $\relvars(e)$ and
every type $\tau$, $\TA \vdash e:\tau$ if and only if there is an
instantiation $\inst$ of $\Gamma$ such that $\TA$ is the image of $\Gamma$
under $\inst$, and such that $\tau$ is the output type of the type formula
under $\inst$.
\end{defin}

The main result of this paper can now succinctly stated as follows:
\begin{theor}[Type inference] \label{theormain}
For every relational algebra expression $e$, there exists a principal
type formula for $e$, which can be effectively
computed from $e$.
\end{theor}
Note that if $e$ is untypable, any unsatisfiable type formula (type formula
with an unsatisfiable type context) is principal for $e$.

We will substantiate our main theorem in the following sections.

\section{Solving systems of set equations} \label{secsolving}

Type inference algorithms for programming languages typically work by
structural induction on program expressions, enforcing the typing rules ``in
reverse,'' and using some form of unification to combine type formulas of
subexpressions.  In our case, relation types are sets,
so we need a replacement for classical unification on terms.  This role will
be played by the following algorithm for solving systems of set equations.

Fix some universe $\UU$.  In principle $\UU$ can be any set, but in our
intended application $\UU$ is the universe of attribute names.  Assume
further given a sufficiently large supply of {\em variables}.  In our
intended application, this role will be played by type variables.

An {\em equation\/} is an expression of the form
${\it lhs} = {\it rhs}$, where both $\it lhs$ and $\it rhs$ are sets of
variables.\footnote{A note on notation:
we will write a set $\{a_1,\ldots,a_n\}$ as
$a_1 \ldots a_n$.}
A {\em system of equations\/} consists of two disjoint sets $L$
and $R$ of variables, and a set of equations, such that
every variable occurring at the left-hand side (right-hand side)
of some equation is in $L$ (in $R$).

A {\em substitution\/} on a set $S$ of variables is a mapping from $S$ to the
subsets of $\UU$.  A substitution is called {\em proper\/} if different
variables are assigned disjoint sets.
A {\em valuation\/}
of a system $\Sigma$ consists of a proper substitution on $L$
and a proper subsitution on $R$.  
A valuation $(f_L,f_R)$ is a {\em solution of\/ $\Sigma$\/} if
for every equation $$ a_1 \ldots a_m = b_1 \ldots b_n $$ in
$\Sigma$, we have $$ f_L(a_1) \cup \cdots \cup f_L(a_m)
= f_R(b_1) \cup \cdots \cup f_R(b_n). $$

A {\em symbolic valuation\/}
of $\Sigma$ consists of a new set $V$ of variables and
a mapping $g$ from $L \cup R$ to the subsets of $V$.
Take some proper substitution $h$ on $V$.
Now define the following substitution $h_L$ on $L$:
for any $a \in L$, $$ h_L(a) := \bigcup \{ h(c) \mid c \in g(a) \}. $$
In a completely analogous way we also
define the substitution $h_R$ on $R$.  We call a symbolic valuation a {\em
symbolic solution of\/ $\Sigma$}
if for every proper substitution $h$ on $V$, the pair
$(h_L,h_R)$ is a solution of $\Sigma$, and conversely, every solution of
$\Sigma$ can be written in this way.  So, a symbolic solution is a finite
representation of the set of all solutions.

As a trivial example, consider the trivial system of equations where
$L=\{a\}$, $R=\{b\}$, and without any equations.  Any valuation is also
a solution.  A symbolic solution is given by $V = \{c_1,c_2,c_3\}$ and
$$ g(a) = c_1 c_2 \qquad {\rm and} \qquad g(b) = c_2 c_3. $$ Indeed,
note that we always work with proper substitutions, so $c_1$, $c_2$ and
$c_3$ stand for pairwise disjoint sets.  In particular, $c_1$ stands for
$a - b$, $c_2$ stands for $a \cap b$, and $c_3$ stands for $b - a$.

\begin{theor} \label{theorequations}
Every system of equations $\Sigma$ has a symbolic solution, which can 
be computed from $\Sigma$ in polynomial time.
\end{theor}
{\bf Proof.}  Let $$ V :=
\{ \bar a \mid a \in L \} \cup \{ \bar b \mid b \in R \}
\cup
\{ (\bar a,\bar b) \mid (a,b) \in (L \times R)\}, $$
and define
the following symbolic valuation $g$ with $V$ as its set of variables:
for each $a \in L$, $$ g(a) := \{ (\bar a,\bar b) \mid b \in R \} \cup \{\bar
a\} $$ and
for each $b \in R$, $$ g(b) := \{ (\bar a,\bar b) \mid a \in L \} \cup \{\bar
b\}. $$
Then define the subset $V_0 \subseteq V$ as follows.  An element $c \in V$ is
in $V_0$ if there is an equation $$ a_1 \ldots a_m = b_1 \ldots b_n $$ in
$\Sigma$ such that $c$ belongs to one of the following two sets but not to the
other: $$ \bigcup_{i=1}^m g(a_i) \qquad {\rm and} \qquad
\bigcup_{j=1}^n g(b_j). $$

Now consider the symbolic valuation $g'$ with $V' := V - V_0$
as its set of variables, defined by
$g'(x) := g(x) - V_0$. This $g'$ can easily be constructed in polynomial time.
We next show that $g'$ is indeed a
symbolic solution of $\Sigma$.

Let $h$ be a proper substitution on $V'$, and let $a_1 \ldots a_m =
b_1 \ldots b_n$ be an equation. By definition of $g'$, for every $i \in
\{1,\ldots,m\}$ and every $c \in g'(a_i)$, there is a $j \in \{1,\ldots,n\}$
such that $c \in g'(b_j)$, and vice versa, for every $j \in \{1,\ldots,n\}$
and every $c \in g'(b_j)$, there is an $i \in \{1,\ldots,n\}$ such that $c \in
g'(a_i)$.  Hence, $$ \bigcup_{i=1}^m
\underbrace{\bigcup \{h(c) \mid c \in g'(a_i)\}}_{\textstyle h_L(a_i)} =
\bigcup_{j=1}^m \underbrace{\bigcup \{h(c) \mid c \in g'(b_j)\}}_{\textstyle
h_R(b_j)} $$
and thus $(h_L,h_R)$ is a solution of $\Sigma$.

Conversely, let $(f_L,f_R)$ be a solution of $\Sigma$. Then define the
following proper valuation $h$ on $V$:
for $a \in L$, $$ h(\bar a) := f_L(a) - \bigcup f_R(R); $$
for $b \in R$, $$ h(\bar b) := f_R(b) - \bigcup f_L(L); $$
and for $(a,b) \in (L \times R)$, $$ h(\bar a,\bar b) :=
f_L(a) \cap f_R(b). $$
Clearly, for each $a \in L$, $$ f_L(a) = \bigcup \{ h(\bar a,\bar b) \mid b
\in R \} \cup h(\bar a), $$ and for each $b \in R$, $$ f_R(b) = \bigcup \{
h(\bar a,\bar b) \mid a \in L \} \cup h(\bar b). $$
Put differently, $$ f_L(a) = \bigcup \{ h(c) \mid c \in g(a) \} $$ for each
$a$, and $$ f_R(b) = \bigcup \{ h(c) \mid c \in g(b) \} $$ for each $b$.
Since we want to show that $g'$ is a symbolic solution, we
would like to show the last two equalities with $g'$ instead of $g$.
Since $g'(x) = g(x) - V_0$,
it suffices to show that $h(c)$ is empty for each $c \in V_0$,

To see that these sets are indeed empty, we consider the three possibilities
for an element of $V$ to be in $V_0$.  If $\bar a \in V_0$ with $a \in L$,
this means that there is some equation $$ a_1 \ldots a_m = b_1 \ldots b_n $$
where $a$ is one of the $a_1$, \ldots, $a_m$.
Since $(f_L,f_R)$ is a solution, $$ f_L(a) \subseteq \bigcup_{j=1}^m f_R(b),
$$ so in particular, since $h(\bar a) \subseteq f_L(a)$,
$$ h(\bar a) \subseteq \bigcup_{j=1}^m f_R(b). $$ However, by definition of
$h$, $h(\bar a)$ is disjoint from each $f_R(b)$. Hence, $h(\bar a)$ must be
empty.

Analogously we see that if $\bar b \in V_0$ with $b \in R$, then $h(\bar b)$
is empty.

So finally, assume $(\bar a,\bar b) \in V_0$ with $(a,b) \in (L\times R)$.
This means that there is either an equation of the form $$ \ldots a \ldots =
\ldots $$ with $b$ not occurring in the right-hand side, or of the form $$
\ldots = \ldots b \ldots $$ with $a$ not occurring in the left-hand side.
Let us focus on the first possibility (the second is analogous) and write the
equation in more detail as $$ \ldots a \ldots = b_1 \ldots b_m. $$
Since $(f_L,f_R)$ is a solution, $f_L(a)$, and in particular $f_L(a) \cap
f_R(b)$, is contained in $\bigcup_{j=1}^m f_R(b_j)$.
However, since $b$ is not among $b_1$, \ldots, $b_m$, and each $f_R(b_j)$ 
is disjoint from $f_R(b)$, this can only be if $f_L(a) \cap f_R(b)$, which is
the same as $h(\bar a,\bar b)$, is empty.
\qed

\bigskip
Let us see a worked-out example of this solution method.
Consider $\Sigma$ with $L = \{a_1,a_2,a_3\}$, $R =
\{b_1,b_2,b_3\}$, and the equations $$ a_1 = b_1 \qquad {\rm and} \qquad a_2 =
b_1b_2. $$ From the first equation we deduce that
$$ \bar a_1, \ (\bar a_1,\bar b_2), \ (\bar a_1,\bar b_3) $$
as well as
$$ \bar b_1, \ (\bar a_2,\bar b_1), \ (\bar a_3,\bar b_1) $$
are in $V_0$. From the second equation we deduce that
$$ \bar a_2, \ (\bar a_2,\bar b_3) $$
as well as
$$ (\bar a_1,\bar b_1), \ \bar b_2, \ (\bar a_3,\bar b_2) $$
are also in $V_0$.  So $$ V - V_0 = \{\bar a_3, \bar b_3, (\bar
a_2,\bar b_2), (\bar a_3,\bar b_3)\}, $$ and the symbolic solution $g'$ is
given by
$$ \begin{array}{ll}
g'(a_1) = \emptyset & g'(b_1) = \emptyset \\
g'(a_2) = (\bar a_2,\bar b_2) & g'(b_2) = (\bar a_2,\bar b_2) \\
g'(a_3) = \bar a_3,\,(\bar a_3,\bar b_3) & g'(b_3) = \bar b_3,\,
(\bar a_3,\bar b_3).
\end{array} $$
If we rename the variables for added clarity,
we obtain the symbolic solution $$ \begin{array}{ll} a_1 = \emptyset & b_1 =
\emptyset \\ a_2 = c_1 & b_2 = c_1 \\
a_3 = c_2c_3 & b_3 = c_3c_4 \end{array} $$
which can be interpreted as specifying that the only solutions to $\Sigma$ are
those where we assign the same set to $a_2$ and $b_2$, which is disjoint from
the sets assigned to $a_3$ and $b_3$ (the latter two sets need not be
disjoint), and where $a_1$ and $b_1$ are empty.

\section{Principal type inference algorithm}

We are now ready to describe our algorithm.  A
computer implementation
is available from the authors \cite{vansummeren}.

\subsection{Two subroutines}

\subsubsection{Extending a type formula with extra special attributes}
\label{secext}

The
following construction will be used as a subroutine in our algorithm.
Let $(\Gamma,\allowbreak e,\allowbreak \outvars,\allowbreak \outatt)$
be a type formula, and let
$A$ be an attribute name not in $\specattrs$.  By {\em extending this type
formula with $A$}, we mean the following:
\begin{enumerate}
\item
add $A$ to $\specattrs$;
\item
define $\constraint(A)$ as
$$ (\bigvee_r r) \to
\bigvee_{a \in \typevars} ( \bigwedge_{a \in \decl(r)} r \land
\bigwedge_{a \not \in \decl(r)} \neg r ); $$
\item
define $\outatt(A)$ as $$ \bigvee \{r \mid \decl(r) \subseteq \outvars\}. $$
\end{enumerate}

\subsubsection{Conjugating two type contexts.} \label{secconjunction}

This is another subroutine that will
be used.  Two type contexts $\Gamma_1$ and $\Gamma_2$ are called {\em
compatible\/} if {\em (i)\/}~$\typevars_1=\typevars_2$; {\em (ii)\/}~$\decl_1$
and $\decl_2$ agree on $\relvars_1 \cap \relvars_2$; and {\em
(iii)\/}~$\specattrs_1=\specattrs_2$.
By the {\em conjunction\/} of two compatible type contexts
$\Gamma_1$ and $\Gamma_2$, we mean the type context defined as follows:
\begin{enumerate}
\item
$\relvars := \relvars_1 \cup \relvars_2$.
\item
$\typevars := \typevars_1$ ($=\typevars_2$).
\item
$\decl := \decl_1 \cup \decl_2$.
\item
$\specattrs := \specattrs_1$ ($=\specattrs_2$).
\item
for each $A \in \specattrs$,
$$ \constraint(A) := \constraint_1(A) \land \constraint_2(A). $$
\end{enumerate}

\subsection{The algorithm}

\subsubsection{Base case} Our algorithm
proceeds by induction on the structure of the expression.  The base case,
where $e$ is a relation variable $r$, is trivial:
$$ r:a \mapsto r:a. $$

\subsubsection{Union}
Let $e=(e_1 \cup e_2)$.  By induction, for
$i=1,2$, we have principal type formulas
$(\Gamma_i,\allowbreak e_i,\allowbreak \outvars_i,\allowbreak \outatt_i)$.
We may assume that $\typevars_1$ and $\typevars_2$ are disjoint.
We perform the following steps:
\begin{enumerate}
\item
For each $A$ in $\specattrs_1$ not in $\specattrs_2$, extend the type formula
for $e_2$ by $A$.
Conversely, for each $A$ in $\specattrs_2$ not in $\specattrs_1$,
extend the type formula for $e_1$ by $A$.  We now have
$\specattrs_1=\specattrs_2$, which we denote by $\specattrs$.
\item \label{equationstep}
Now consider the system of set equations $\Sigma$ with $L = \typevars_1$, $R
= \typevars_2$, and the set of equations 
$$ \displaylines{ \{ \decl_1(r) = \decl_2(r) \mid r \in \relvars_1 \cap
\relvars_2 \} \hfill \cr \hfill {} \cup \{
\outvars_1 = \outvars_2 \}. \cr} $$
Find a symbolic solution to this system, and
apply it to the two type formulas.  Denote the result of applying the solution
to $\outvars_1$ by $\outvars$; by the equation $\outvars_1=\outvars_2$, this
is the same as the result of applying the solution to $\outvars_2$.
\item
The two type contexts
$\Gamma_1$ and $\Gamma_2$ have now become compatible; in particular, they have
the same set of type variables, which we denote by $\typevars$.
Take their conjunction $\Gamma$.   The resulting set of relation variables is
denoted by $\relvars$.  The resulting constraint mapping is denoted by
$\constraint'$.
\item \label{outstep}
For each $A$ in $\specattrs$, define $\constraint(A)$ as $$ \constraint'(A)
\land (\outatt_1(A) \leftrightarrow \outatt_2(A)), $$ and define $\outatt(A)$
as $\outatt_1(A)$.
\end{enumerate}
The result is a principal type formula $(\Gamma,\allowbreak e,\allowbreak
\outvars,\allowbreak \outatt)$ for $e$.

\subsubsection{Difference}
The case $e=(e_1 - e_2)$ is
treated in exactly the same way as the case $e=(e_1 \cup e_2)$.

\subsubsection{Natural join}
The case $e=(e_1 \Join e_2)$ is treated as the case
$e=(e_1 \cup e_2)$, except for the following important differences in two of
the steps:
\begin{enumerate}
\item[\ref{equationstep}.]
We omit the equation $\outvars_1 = \outvars_2$ from the system of equations.
We now define $\outvars$ as the union of the results of applying the symbolic
solution to $\outvars_1$ and $\outvars_2$.
\item[\ref{outstep}.]
For each $A$ in $\specattrs$, $\constraint(A)$ is now the same as
$\constraint'(A)$, and $\outatt(A)$ is now defined as $$ \outatt_1(A) \lor
\outatt_2(A). $$
\end{enumerate}

\subsubsection{Cartesian product}
The case $e=(e_1 \times e_2)$ is treated as the
case $e=(e_1 \Join e_2)$, except for the following two differences, again in
steps \ref{equationstep} and \ref{outstep}:
\begin{enumerate}
\item[\ref{equationstep}.]
In the computation of the symbolic
solution, we put every pair $(\bar a,\bar b)$
with $a \in \outvars_1$ and $b \in
\outvars_2$ by default in $V_0$ (cf.~the solution method
described in the proof of Theorem~\ref{theorequations}).  This will guarantee
that the results of applying the solution to $\outvars_1$ and $\outvars_2$
will be disjoint.
\item[\ref{outstep}.]
For each $A$ in $\specattrs$, define $\constraint(A)$ as $$ \constraint'(A)
\land \neg (\outatt_1(A) \land \outatt_2(A)). $$
\end{enumerate}

\subsubsection{Selection}
Let $e = \sigma_{\theta(A_1,\ldots,A_n)}(e')$.
\begin{enumerate}
\item
Initialize the desired type formula $$ (\Gamma,\allowbreak e,\allowbreak
\outvars,\allowbreak \outatt) $$ to
the principal type formula $(\Gamma',\allowbreak e',\allowbreak
\outvars',\allowbreak \outatt')$ for $e'$
(which we already have by induction).
\item
For $i=1,\ldots,n$,
if $A_i$ is not yet in $\specattrs$, extend the type formula with $A_i$.
\item \label{forcestepin}
for $i=1,\ldots,n$,
replace $\constraint(A_i)$ by $$ \constraint(A_i) \land \outatt(A_i). $$
\item \label{forcestepout}
For $i=1,\ldots,n$, put $\outatt(A_i):={\bf true}$.
\end{enumerate}

\subsubsection{Projection}
For the case $e = \pi_{A_1,\ldots,A_n}(e')$
we do the same as for
the case $e = \sigma_{\theta(A_1,\ldots,A_n)}(e')$.
In addition, we set
\begin{itemize}
\item
$\outatt(A) := {\bf false}$ for each $A$ in $$ \specattrs -
\{A_1,\ldots,A_n\}, $$ and 
\item
$\outvars := \emptyset$.
\end{itemize}

\subsubsection{Renaming}
The case $e = \rho_{A/B}(e')$ is treated similarly
to the case $e = \sigma_{\theta(A,B)}(e')$, except
that we treat $B$ differently from $A$
in step~\ref{forcestepin}, as follows:
\begin{enumerate}
\item[\ref{forcestepin}.] Replace $\constraint(B)$ by $$ \constraint(B) \land
\neg \outatt(B). $$
\end{enumerate}
Furthermore, step~\ref{forcestepout} is changed as follows:
\begin{enumerate}
\item[\ref{forcestepout}.] Put $\outatt(A):={\bf false}$, and
$\outatt(B):={\bf true}$.
\end{enumerate}

\subsubsection{Projecting out}
Finally, the case $e = \widehat \pi_A(e')$ is
treated similarly to $e = \sigma_{\theta(A)}(e')$, with the exception
that we set $\outatt(A):={\bf false}$ instead of ${\bf true}$.

\subsection{Example}

We illustrate
the working of our algorithm on the
expression $$ e = \sigma_{B=C} \underbrace{( \underbrace{(
\underbrace{\rho_{A/B}(r)}_{e_1} \cup
s )}_{e_2} \Join u )}_{e_3}. $$
We will encounter only rather trivial
systems of equations in doing this
example; the reader is invited to try the example expression discussed in the
Introduction for more interesting systems of equations.

To find the type formula for $e_1$, we start
from the trival type formula $r:a \mapsto r:a$ for $r$.  Extending this type
formula with $A$ and $B$ yields
$$ \begin{array}{lcl} r:a & \mapsto & r:a \\
A:r\to r && A:r \\ B:r\to r && B:r. \end{array} $$
Then we change the constraint $r\to r$ (or simply $\bf true$)
for $A$ by ${\bf true} \land r$, or
simply $r$, and we 
change the constraint for $B$ by ${\bf true} \land \neg r$, or $\neg r$.
Finally, we set $\outatt(A)$ to $\bf false$ and
$\outatt(B)$ to $\bf true$, yielding:
$$ \begin{array}{lcl} r:a & \mapsto & e_1:a \\
A:r && A:{\bf false} \\ B:\neg r && B:{\bf true}. \end{array} $$

To find the type formula for $e_2$, we start from that for $e_1$ and the
trivial formula for $s$, which we extend with $A$ and $B$ as
$$ \begin{array}{lcl} s:b & \mapsto & s:b \\
A:{\bf true} && A:s \\ B:{\bf true} && B:s. \end{array} $$
We now consider the rather trivial system of set equations with $L=\{a\}$,
$R=\{b\}$, and the single equation $a=b$.  The symbolic solution is obviously
$a=c, b=c$. Applying this solution to the two type formulas simply changes
both $a$ and $b$ into $c$.  Conjugating the two type contexts yields the
constraint
$r \land {\bf true}$ for $A$, which can be
simplified to $r$, and the constraint $\neg r \land {\bf true}$ for $B$,
which can be simplified to $\neg r$.
Then we add the conjunct ${\bf false} \leftrightarrow s$ to
the constraint for $A$, yielding $r \land \neg s$, and we add the conjunct
${\bf true} \leftrightarrow s$ for $B$, yielding $\neg r \land s$.  Finally,
$\outatt(A)$ is set to $\bf false$, and $\outatt(B)$ to $\bf true$, yielding:
$$ \begin{array}{lcl} \begin{array}{l} r:c \\ s:c \end{array} & \mapsto &
e_2:c \\ A:r \land \neg s && A:{\bf false} \\ B:s \land \neg r && B:{\bf true}.
\end{array} $$

To find the type formula for $e_3$, we start from the one for $e_2$ and the
trivial formula for $u$, which we extend with $A$ and $B$ as
$$ \begin{array}{lcl} u:d & \mapsto & u:d \\ A:{\bf true} && A:u \\
B:{\bf true} && B:u. \end{array} $$
We now get the even more trivial
system of set equations with $L=\{c\}$, $R=\{d\}$, and no equations, which has
as symbolic solution $c=c_1c_2, d=c_2c_3$. We set $\outvars$ to $c_1c_2c_3$.
Conjugating the two type contexts (after having filled in the solution) yields
nothing surprising.  Finally we
set $\outatt(A)$ to ${\bf false} \lor u$, which
simplifies to $u$, and set $\outatt(B)$ to ${\bf true} \lor u$, or simply $\bf
true$, yielding:
$$ \begin{array}{lcl} \begin{array}{l} r:c_1c_2 \\ s:c_1c_2 \\ u:c_2c_3
\end{array} & \mapsto &
e_3:c_1c_2c_3 \\
A:r \land \neg s && A:u \\ B:s \land \neg r && B:{\bf true}.
\end{array} $$

Finally, to find the type formula for $e$ itself, we first extend the one for
$e_3$ with $C$:
$$ \begin{array}{lcl} \begin{array}{l} r:c_1c_2 \\ s:c_1c_2 \\ u:c_2c_3
\end{array} & \mapsto &
e_3:c_1c_2c_3 \\
A:r \land \neg s && A:u \\ B:s \land \neg r && B:{\bf true} \\
C:\varphi && C:r \lor s \lor u.
\end{array} $$
Here, $\varphi$ is the formula 
$$ (r \lor s \lor u) \to ((r \land s \land \neg u) \lor (r \land s \land u)
\lor (\neg r \land \neg s \land u)), $$ or simply $r \leftrightarrow s$.
Then we add the conjunct $\bf true$ to the constraint for $B$ (which has no
effect), and the conjunct $(r \lor s \lor u)$ to the constraint for $C$.
Finally, we set $\outatt(B) = \outatt(C) = {\bf true}$, yielding indeed
the type
formula we gave for $e$ in Section~\ref{secexamples} (modulo renaming of type
variables).

\subsection{Correctness proof}

Extension of a type formula with extra special attributes
(Section~\ref{secext})
is a heavily used subroutine
in our type inference algorithm, and one might even go as far as saying that
it is the only part of the algorithm whose correctness is not self-evident.
Hence, the following lemma is of crucial importance:

\begin{lemma}
\label{mainlemma}
The extension of any type formula, generated by our algorithm,
with an extra special attribute, always produces an equivalent type formula.
\end{lemma}
Here, equivalence naturally means the following.
Consider two type formulas $\Phi_1$ and $\Phi_2$ whose type contexts
$\Gamma_1$ and $\Gamma_2$ have the same $\relvars$, and let $\inst_1$
($\inst_2$) be an instantiation of $\Gamma_1$ ($\Gamma_2$). We say that
{\em $\inst_1$ and $\inst_2$ are equivalent with respect to
$\Phi_1$ and $\Phi_2$\/} if $\inst_1(\Gamma_1) =
\inst_2(\Gamma_2)$, and the output type of $\Phi_1$ under $\inst_1$
equals the output type of $\Phi_2$ under $\inst_2$. We say that {\em
$\Phi_1$ and $\Phi_2$ are equivalent\/} if for every instantiation of
$\Gamma_1$ there is an equivalent instantiation of $\Gamma_2$, and vice
versa.

Now to the proof of Lemma~\ref{mainlemma}.
Let $\Phi=(\Gamma,e,\outatt,\outvars)$ be a type formula, and let $\Phi' =
(\Gamma',\allowbreak e,\allowbreak \outatt',\allowbreak \outvars)$
be its extension with the extra special
attribute $A$.  We have to show that $\Phi$ and $\Phi'$ are equivalent.

\paragraph{From $\Phi$ to $\Phi'$.}
Let $\inst$ be an instantiation of $\Gamma$.  We have to find an
equivalent instantiation $\inst'$ of $\Gamma'$.

If $A \not \in \inst(a)$ for every $a \in \typevars$, we can simply put
$\inst'(a):=\inst(a)$ for each type variable $a$, $\inst'(B):=\inst(B)$ for
each special attribute $B \neq A$, and $\inst'(A):=\emptyset$.
In this case, it is clear that $\inst'$ is a legal instantiation of $\Gamma'$,
that $\inst(\Gamma)=\inst'(\Gamma')$, and that the output type of $\Phi$ under
$\inst$ equals the output type of $\Phi'$ under $\inst'$.

If $A \in \inst(a)$ for some $a \in \typevars$, we put $\inst'(a) :=
\inst(a)-\{A\}$ for this $a$, and put $\inst'(b):=\inst(b)$ for every type
variable $b \neq a$.  We also put $\inst'(B):=\inst(B)$ for each special
attribute $B \neq A$.  We finally put $\inst'(A) := \{ r \mid a \in \decl(r)
\}$.
It is clear that $\inst'$ is a legal instantiation of $\Gamma'$, and that
$\inst(\Gamma)=\inst'(\Gamma')$.  To show that the output type of $\Phi$ under
$\inst$ equals the output type of $\Phi'$ under $\inst'$, we must show that
if $a \in \outvars$, then there exists an
$r \in \inst'(A)$ such that $\decl(r) \subseteq \outvars$.  We will do this
in Lemma~\ref{usefullemma}.

\paragraph{From $\Phi'$ to $\Phi$.}
Let $\inst'$ be an instantiation of $\Gamma'$.  We have to find an
equivalent instantiation $\inst$ of $\Gamma$.

If $\inst'(A)=\emptyset$, then we put $\inst(a):=\inst'(a)$ for each type
variable $a$, and $\inst(B):=\inst'(B)$ for each special attribute $B \neq A$.
In this case it is clear that $\inst'(\Gamma)=\inst(\Gamma)$, and that the
output type of $\Phi'$ under $\inst'$ equals the output type of $\Phi$ under
$\inst$.

If $\inst'(A)\neq\emptyset$, we know (because
$\inst'(A) \models \constraint'(A)$) that there exists an $a \in \typevars$
such that $\inst'(A) = \{r \mid a \in \decl(r)\}$.
Then we put $\inst(a):=\inst'(a) \cup \{A\}$, and $\inst(b) :=
\inst'(b)$ for each type variable $b \neq a$.  We also put $\inst(B) :=
\inst'(B)$ for each special attribute $B \neq A$.
It is now again clear that $\inst'(\Gamma') = \inst(\Gamma)$, and
that the output type of $\Phi'$ under
$\inst'$ equals the output type of $\Phi$ under $\inst$.
\qed

We still owe:

\begin{lemma}
\label{usefullemma}
In any type formula generated by our algorithm, the following
holds.  Let $a$ be a type variable in $\outvars$.  Then there exists a
relation
variable $r$ such that $a \in \decl(r)$ and $\decl(r) \subseteq \outvars$.
\end{lemma}
{\bf Proof.} By induction.  The base case, $r:a \mapsto r:a$, is trivial.

For the case $e = (e_1 \cup e_2)$ we reason as follows.  Let $g$ be the
symbolic solution to the system of equations.  Then $\outvars = \bigcup
g(\outvars_1) = \bigcup g(\outvars_2)$.  Let $c \in \outvars$.  Then $c
\in g(a)$ for some $a \in \outvars_1$.  By induction, we know that for some
relation variable $r$, $a \in \decl_1(r)$ and $\decl_1(r) \subseteq
\outvars_1$.  This implies that $c \in \bigcup g(\decl_1(r)) = \decl(r)$, and
that $\decl(r) \subseteq \outvars$.

For the case $e = (e_1 \Join e_2)$ we have $\outvars$ equal to
$\bigcup g(\outvars_1)
\cup \bigcup g(\outvars_2)$, $g$ again being the symbolic solution.  Let $c
\in \outvars$.  So, $c \in \bigcup g(\outvars_1)$ or $c \in \bigcup
g(\outvars_2)$.  By symmetry we may assume that $c \in \bigcup g(\outvars_1)$.
Then $c \in g(a)$ for some $a \in \outvars_1$.  By induction, we know that for
some $r$, $a \in \decl_1(r)$ and $\decl_1(r) \subseteq \outvars_1$.  This
implies again that $c \in \decl(r)$ and $\decl(r) \subseteq \outvars$.

For the case $e = (e_1 \times e_2)$, we can use exactly the same reasoning as
for $(e_1 \Join e_2)$, because no particular properties of the symbolic
solution have been used.

The cases $e = \sigma$, $\rho$ and $\widehat\pi$ are trivial because they
don't change $\outvars$ and $\decl$.  The case $e = \pi$ is trivial because it
sets $\outvars$ to $\emptyset$.  \qed

\bigskip

By induction on the structure of relational algebra expressions we can now
prove that each case of our algorithm correctly produces a type formula that
is principal.  The cases corresponding to unary operators are all proven
correct in an analogous way; we treat the selection as an example below.
The cases
corresponding to binary operators heavily rely in addition on the correctness
of our algorithm for solving systems of set equations, which we already proved
correct in Section~\ref{secsolving}.

So, let $e = \sigma_{\theta(A_1,\ldots,A_n)}(e')$.
Let the type formulas computed
by our algoritm for $e$ and $e'$ be $\Phi$ and $\Phi'$, respectively.
By induction, we may assume that $\Phi'$ is principal for $e'$; we must show
that $\Phi$ is principal for $e$.

By Lemma~\ref{mainlemma}, we may ignore step~2 of the algorithm and
assume without loss of generality that for $i=1,\ldots,n$, $A_i$ is
already in ${\it Specattrs}'$.  More generally, we may assume that
$\Phi$ differs from $\Phi'$ only in that for $i=1,\ldots,n$, $$ {\it
constraint}(A_i) = {\it constraint}'(A_i) \land {\it outatt}'(A_i) $$
and $$ {\it outatt}(A_i) = {\bf true}. $$

Now suppose ${\cal T} \vdash e:\tau$.  We must find an instantiation $\cal I$
of $\Gamma$ such that $\cal T$ equals ${\cal I}(\Gamma)$ and $\tau$ equals the
output type of $\Phi$ under $\cal I$.  Since ${\cal T} \vdash e:\tau$, we know
that ${\cal T} \vdash e':\tau$ and that for $i=1,\ldots,n$, $A_i \in \tau$.
Since $\Phi'$ is principal for $e'$, we know furthermore that there exists an
instantiation ${\cal I}'$ of $\Gamma'$ such that $\cal T$ equals ${\cal
I}'(\Gamma')$ and $\tau$ equals the output type of $\Phi'$ under ${\cal I}'$.
We set the desired $\cal I$ simply equal to ${\cal I}'$, and verify:
\begin{itemize}
\item
${\cal I}$ is a valid instantiation of $\Gamma$: Thereto, we must check
for $i=1,\ldots,n$ that
${\cal I}(A_i) \models {\it constraint}(A_i)$, or
${\cal I}'(A_i) \models {\it constraint}'(A_i) \land {\it outatt}'(A_i)$,
which is equivalent.
That ${\cal I}'(A_i) \models {\it constraint}'(A_i)$ is trivial, by definition.
That ${\cal I}'(A_i) \models {\it outatt}(A_i)$ is also clear, since $A_i \in
\tau$ and $\tau$ equals the output type of $\Phi'$ under ${\cal I}'$.
\item
${\cal T} = {\cal I}(\Gamma)$: This is clear, since ${\cal T} = {\cal
I}'(\Gamma')$ and ${\cal I}(\Gamma) = {\cal I}'(\Gamma')$.
\item
$\tau$ equals the output type of $\Phi$ under $\cal I$: 
Since $\it outatt$ differs from ${\it outatt}'$ only in that the output 
constraints for the $A_i$ are loosened,
the output type of $\Phi'$ under ${\cal I}'$, which equals $\tau$, can only be
a subset of the output type of $\Phi$ under $\cal I$.  However, as every $A_i$
is already in $\tau$, this subset relationship cannot be a strict one, and
hence the two types are indeed equal.
\end{itemize}

Conversely, suppose $\cal I$ is an instantiation of $\Gamma$, and let $\tau$
be the output type of $\Phi$ under $\cal I$.  We must now show that ${\cal
I}(\Gamma) \vdash e:\tau$.  To show this, we note that $\cal I$ is a valid
instantiation of $\Gamma'$ (as the attribute constraints of $\Gamma$ are
tighter than those of $\Gamma'$).  Hence, since $\Phi'$ is principal for $e'$,
we know that ${\cal I}(\Gamma) = {\cal I}(\Gamma') \vdash e':\tau'$,
where $\tau'$ is the output type of $\Phi'$ under $\cal I$.  But this output
type
is the same as
the output type of $\Phi$ under $\cal I$; indeed, $\it outatt$ differs
only from ${\it outatt}'$ on the $A_i$, but all $A_i$ are members of
both types anyway (for $\Phi'$ this is because ${\cal I}(A_i)$ satisfies ${\it
outatt}'(A_i)$ by definition, and for $\Phi$ this is trivial because ${\it
outatt}(A_i) = {\bf true}$).  Hence, we have ${\cal I}(\Gamma) \vdash e':\tau$.
Since all the $A_i$ are in $\tau$, we can conclude that
${\cal I}(\Gamma) \vdash e:\tau$.

\subsection{Complexity and typability} \label{subsecuntyp}

Since every step of the induction can be implemented in time polynomial
in the size of the output of its child steps, a rough upper bound on the time
complexity of our algorithm is $2^{2^{O(n)}}$.   It remains open whether this
complexity can be improved.
Note that type formulas can be
exponentially large; for example, the type
formula for $r_1 \Join (r_2 \Join (\cdots \Join r_m) \cdots )$ uses
$O(2^m)$ different type variables.

If the input expression was untypable, the algorithm will output an
unsatisfiable type formula.  Hence,
an alternative way to check typability
is to check satisfiability of the principal type formula.
We do not have to wait until the end, however, to test satisfiability.
In principle, as soon as an unsatisfiable
attribute constraint arises during type inference, the algorithm can
stop and report that the expression is untypable.  This is more useful,
since it tells exactly where the expression breaks down.  In a practical
implementation, one could do this by keeping the attribute constraints in
disjunctive normal form.  Doing this might actually have a better complexity
than expected, since the attribute constraints generated by the algorithm have
a quite special form, which might be exploited.

Note that unsatisfiable attribute constraints can only be generated 
in the following places:
\begin{itemize}
\item
Step~\ref{outstep} of cases $\cup$ and $-$, and its adaptation for
case $\times$.  A simple
example of a type error that will be spotted in this place
is $\pi_A(r) \cup \pi_B(s)$.
\item
Step~\ref{forcestepin} of case $\sigma$, and its analogues for
$\pi$, $\rho$, and $\widehat \pi$. A simple example of a type error that will
be spotted in this place is $\sigma_{\theta(A)}(\pi_B(r))$.
\end{itemize}
Since the above-mentioned steps in the algorithm are clearly
only executed if there are special attributes, we thus have:

\begin{propo}
Every expression without special attributes is typable.
\end{propo}
The reader might wonder about contrived examples such as
$$ (r \times s) \Join (r \cup s), $$ which has no special attributes,
but does not seem typable.  However,
this expression is well-typed under the type assignment by which the types of
$r$ and $s$ are empty.

\section{Polymorphic queries}

Usually, a query is defined as a mapping from databases of some fixed
type to relations of some fixed type.  We can define a polymorphic
generalization of the notion of query, to allow databases of different
types as input.  Fix a schema $\Sch$.
\begin{defin} \rm
\begin{enumerate}
\item
Let $\TA$ be a type assignment on $\Sch$, and let $\tau$ be a type.
A {\em query of type $\TA \to \tau$} is a mapping from databases of type $\TA$
to relations of type $\tau$.
\item
An {\em input-output type family\/} is a {\em partial\/} function $F$
from all type assignments on $\Sch$ to all types.
We denote the definition domain of $F$ by $\mathop{\rm dom} F$.
\item
A {\em polymorphic query of type $F$} is a family $(Q_{\TA})_{\TA \in
\mathop{\rm dom} F}$ of queries, where each $Q_{\TA}$ is a query of type $\TA
\to F(\TA)$.
\end{enumerate}
\end{defin}

Viewed from this perspective,
a type formula $\gamma$ with type context $\Gamma$ is,
of course, nothing but a specification of an input-output
family $F_\gamma$:  we have $\mathop {\rm dom} F_\gamma = \{\inst(\Gamma) \mid
\inst$ an instantiation of $\Gamma\}$, and $F_\gamma(\inst(\Gamma))$ equals
the output type of $\gamma$ under $\inst$.
As a consequence, every relational algebra expression $e$
expresses a polymorphic query of type $F_\gamma$, where $\gamma$ is
the principal type formula for $e$.

The following notion now naturally presents itself:
\begin{defin} \rm Two relational algebra expressions $e_1$ and $e_2$ are {\em
polymorphically equivalent\/} if they express the same polymorphic query.
\end{defin}

For example, the equivalence 
$$ \sigma_{A=B}(r \times \pi_{A,B,C}(s)) \equiv
r \times \sigma_{A=B}\pi_{A,B,C}(s) $$ is polymorphic, but the equivalence
$$\pi_A(r \Join \pi_{A,B}(s)) \equiv 
\pi_A(r \Join s)$$ is not, as it is only
valid under a type assignment $\TA$
such that $\TA(r) \cap \TA(s)$ is a subset of $\{A,B\}$.

We are now weaponed to return to the issue of non-redundancy
already touched upon
at the end of Section~\ref{prelims}.
\begin{propo} \label{nonredundant}
\begin{enumerate}
\item
There is no expression not using $\Join$ that is polymorphically equivalent to
$r \Join s$.  We say that $\Join$ is\/ {\em polymorphically non-redundant}.
The same holds for the operator $\times$.
\item
There is no expression not using $\pi$ that is polymorphically equivalent
to $\pi_A(r)$.  So, also $\pi$ is polymorphically non-redundant.  The same
holds for the operator $\widehat \pi$.
\end{enumerate}
\end{propo}
{\bf Proof.}
Any expression $e$ polymorphically equivalent to $r \Join s$
must have principal type $$ \begin{array}{l} r:a_1a_2
\\ s:a_2a_3 \end{array} \mapsto e:a_1a_2a_3. $$  
Inspecting the principal type inference algorithm, we see that a type formula
where $\outvars$ contains the union of $\decl(r)$ and $\decl(s)$, where the
latter two sets are different and have a non-empty intersection, can only be
produced in the case of $\Join$.
An analogous argument deals with $\times$.

As for $\pi_A(r)$, any polymorphically equivalent expression $e$
must have principal type $$ \begin{array}{lcl} r:a & \mapsto & e:\emptyset \\
A:r && A:{\bf true} \end{array}. $$  Inspecting the principal type inference
algorithm, we see that a type formula where $\outvars$ is made
empty, depending on some special attribute, can only be
produced in the case of $\pi$.
An analogous argument deals with $\widehat \pi$. \qed

\bigskip
We can also show polymorphic inexpressibility results for the full language.
For example:
\begin{propo} \label{semiprimitive}
The semijoin $r \semijoin s$
is not\/ {\em polymorphically\/} expressible in the standard
relational algebra.
\end{propo}
{\bf Proof.}
Suppose $e$ is an expression polymorphically equivalent to $r \semijoin s$.
The principal type of $e$ must be
$$ \begin{array}{l} r : a_1a_2 \\ s: a_2a_3
\end{array} \mapsto e : a_1a_2. $$  Since there are no special attributes, the
operators $\sigma$, $\rho$, $\widehat \pi$, and $\pi$ cannot occur in $e$,
except for $\pi_\emptyset$ (projection on the empty sequence of attributes).
Now consider the type assignment $\TA$ on $\{r,s\}$ given by $\TA(r) =
\{A,B\}$ and $\TA(s) = \{B,C\}$, and the
database $\DB$ of type $\TA$ defined by $\DB(r) =
\{[A:x,\allowbreak B:y],\allowbreak [A:u,\allowbreak B:v]\}$ and
$\DB(s) = \{[B:y,\allowbreak C:z]\}$.  Given $\TA$, the type
of $e$ is $\{A,B\}$. Using the above knowledge of $e$,
we can see that in the value of $e$ on $\DB$, either $[A:x,\allowbreak
B:y]$ and $[A:u,\allowbreak B:v]$ both occur,
or none of them occurs.  However, this is in contradiction
with the fact that $e$ is equivalent to $r \semijoin s$.  Hence, $e$ does not
exist.
\qed

\section{Concluding remarks}

We have seen in the previous section that
classical ``derived'' operators of the standard relational algebra
can become primitive in the polymorphic setting.  The same holds for many other
such operators. Note that
it is actually easy to extend our type inference
algorithm to include semijoin and similar operators, so
Proposition~\ref{semiprimitive}
should not be misinterpreted as a negative result.  Rather, it
indicates that the new issue arises as to how a basic polymorphic query
language should be designed.  This is an interesting direction for further
work.

As already mentioned in the Introduction, other obvious directions for further
work 
include (i) applying type inference in practice to SQL rather than
to the relational algebra; (ii) developing type inference in the context
of semi-structured data models rather than the relational data model; or
(iii) to do the same for object-oriented query languages such as OQL\@.
When moving to the OO context, one has to deal with the additional
subtilities created by inheritance and subtyping.  Current research
in programming languages is giving these issues considerable attention.

We have also ignored types on the level of individual attribute values,
although such types are almost always present in practice, e.g., in
SQL\@.  For example, for $\sigma_{A = \text{``John''}}(r)$ to be
well-typed it suffices for us that the type of $r$ has an
$A$-attribute.  However, in reality, $A$ must in addition be of type
string.  Incorporating types on the attribute value level only has an
effect on the special attributes of an expression; it has no effect on
its polymorphic basis (recall the notion of polymorphic basis from
Section~\ref{secexamples}).  Hence, a type inference algorithm can still be
based on solving systems of set equations.  When conjugating two type
contexts, however (recall Section~\ref{secconjunction}),
a unification on the value types associated to the
special attributes has to be performed.  A similar unification is
induced by the natural join operator.  Moreover, in the case of the
selection operator, the selection predicate (which in our approach has
remained abstract) will perform certain operations on certain special
attributes, which will induce certain constraints on the value types
associated to these attributes.  In general, if
the programming language in which we write selection
predicates has a unification-based type system, then we can simply activate
type inference for this system at the appropriate places.

\section*{Acknowledgment}

We thank Serge Abiteboul, who suggested the idea of type inference for
relational algebra to the second author many years ago; Didier R\'emy
and Limsoon Wong, for helpful conversations;
and Julien Forest and Veronique Fischer, who implemented preliminary
versions of the algorithm.


\begin{thebibliography}{10}

\bibitem{asu}
A.V. Aho, R.~Sethi, and J.D. Ullman.
\newblock {\em Compilers}.
\newblock Addison-Wesley, 1986.

\bibitem{buneman_structure}
P.~Buneman, S.~Davidson, M.~Fernandez, and D.~Suciu.
\newblock Adding structure to unstructured data.
\newblock In F.~Afrati and Ph. Kolaitis, editors, {\em Database
  Theory---ICDT'97}, volume 1186 of {\em Lecture Notes in Computer Science},
  pages 336--350. Springer, 1997.

\bibitem{unql}
P.~Buneman, S.~Davidson, G.~Hillebrand, and D.~Suciu.
\newblock A query language and optimization techniques for unstructured data.
\newblock In {\em Proceedings of the 1996 {ACM SIGMOD} International Conference
  on Management of Data}, issue 25:2 of {\em SIGMOD Record}, pages 505--516.
  ACM Press, 1996.

\bibitem{tsimmis}
H.~Garcia-Molina, Y.~Papakonstantinou, D.~Quass, A.~Rajaraman, Y.~Sagiv,
  J.~Ullman, V.~Vassalos, and J.~Widom.
\newblock The {TSIMMIS} approach to mediation: data models and languages.
\newblock {\em Journal of Intelligent Information Systems}, 8(2):117--132,
  1997.

\bibitem{giannini_survey}
P.~Giannini, F.~Honsell, and S.~Ronchi~della Rocca.
\newblock Type inference: some results, some problems.
\newblock {\em Fundamenta Informaticae}, 19:87--125, 1993.

\bibitem{guntermitchell_taoop}
C.A. Gunter and J.C. Mitchell, editors.
\newblock {\em Theoretical Aspects of Object-Oriented Programming}.
\newblock MIT Press, 1994.

\bibitem{hindley}
J.R. Hindley.
\newblock {\em Basic Simple Type Theory}.
\newblock Cambridge University Press, 1997.

\bibitem{melton}
J.~Melton.
\newblock {\em Understanding {SQL}'s Stored Procedures}.
\newblock Morgan Kaufmann, 1998.

\bibitem{mitchell}
J.C. Mitchell.
\newblock {\em Foundations for Programming Languages}.
\newblock MIT Press, 1996.

\bibitem{machiavellij}
A.~Ohori and P.~Buneman.
\newblock Polymorphism and type inference in database programming.
\newblock {\em ACM Transactions on Database Systems}, 21(1):30--76, 1996.

\bibitem{machiavellic}
A.~Ohori, P.~Buneman, and V.~Breazu-Tannen.
\newblock Database programming in {M}achiavelli---a polymorphic language with
  static type inference.
\newblock In {\em Proceedings
  of the 1989 {ACM SIGMOD} International Conference on the Management of Data},
  issue 18:2 of {\em SIGMOD Record}, pages 46--57. ACM Press, 1989.

\bibitem{remy_records}
D.~R\'emy.
\newblock Type inference for records in a natural extension of {ML}.
\newblock In Gunter and Mitchell \cite{guntermitchell_taoop}, pages 67--96.

\bibitem{remy_concatenation}
D.~R\'emy.
\newblock Typing record concatenation for free.
\newblock In Gunter and Mitchell \cite{guntermitchell_taoop}, pages 351--372.

\bibitem{stemple_poly}
D.~Stemple et~al.
\newblock Exceeding the limits of polymorphism in database programming
  languages.
\newblock In F.~Bancilhon, C.~Thanos, and D.~Tsichritzis, editors, {\em
  Advances in Database Technology---EDBT'90}, volume 416 of {\em Lecture Notes
  in Computer Science}, pages 269--285. Springer-Verlag, 1990.

\bibitem{tiuryn_survey}
J.~Tiuryn.
\newblock Type inference problems: a survey.
\newblock In B.~Rovan, editor, {\em Mathematical Foundations of Computer
  Science}, volume 452 of {\em Lecture Notes in Computer Science}, pages
  105--120, 1990.

\bibitem{ullman_ml}
J.D. Ullman.
\newblock {\em Elements of ML Programming}.
\newblock Prentice-Hall, 1998.

\bibitem{vansummeren}
S. Vansummeren. An implementation of polymorphic type inference for the
relational algebra, written in the programming language ML\@. Master's thesis,
University of Maastricht, 2001.

\end{thebibliography}
\end{document}